\documentclass[aps,prb,twocolumn,floatfix]{revtex4}
\usepackage{graphicx}
\usepackage{bm}
\usepackage{color}
\DeclareGraphicsExtensions{.pdf,.eps,.png,.jpg,.mps}
\begin{document}
\title{Effective approach to the Nagaoka regime of the two dimensional $t$--$J$ model}
\author{M. M. Ma{\'s}ka}
\affiliation{Department of Theoretical Physics,
Institute of Physics, University of
Silesia, 40--007 Katowice, Poland}
\author{M. Mierzejewski}
\affiliation{Department of Theoretical Physics, Institute of Physics,
University of Silesia, 40--007 Katowice, Poland}
\author{E. A. Kochetov}
\affiliation{Theoretical Physics Laboratory, Joint Institute for
Nuclear Research, 141980 Dubna, Russia}
\author{L. Vidmar}
\affiliation{ J. Stefan Institute, SI-1000 Ljubljana, Slovenia }
\author{J. Bon\v{c}a}
\affiliation{ J. Stefan Institute, SI-1000 Ljubljana, Slovenia }
\affiliation{Department of Physics, FMF, University of Ljubljana, Jadranska 19, SI-1000 Ljubljana, Slovenia }
\author{O. P. Sushkov}
\affiliation{School of Physics, University of New South Wales, 2052 Sydney, Australia}

\begin{abstract}
We argue that the $t$--$J$ model and the recently proposed Ising
version of this model give the same physical picture
of the Nagaoka regime for  $J/t \ll 1$. In particular,
both models are shown to give compatible results for a single Nagaoka polaron
as well as for a Nagaoka bipolaron. When compared to the standard $t$--$J$
or  $t$--$J_z$ models, the Ising version allows for a numerical analysis on much
larger clusters by means of classical Monte Carlo simulations. Taking the advantage
of this fact, we study the low doping regime of  $t$--$J$ model for $J/t \ll 1$ and
show that the ground state exhibits phase separation into hole--rich ferromagnetic (FM)
and hole--depleted antiferromagnetic (AFM) regions.
This picture holds true up to a threshold concentration of holes, $\delta \le \delta_t \simeq 0.44\sqrt{J/t}$. Analytical calculations show that $\delta_t=\sqrt{J/ 2\pi t}$.
\end{abstract}
\maketitle

\section{Introduction}

It is by and large  recognized  that the key properties of high-temperature superconducting materials
can be explained with the help of the two-dimensional $t$--$J$ model.\cite{spalek,pwa,zr} Therefore,
the major interest is
focused on this model in the regime that seems to be relevant to high-temperature superconductivity.
Since the Hubbard on-site repulsion $U$ in cuprates is
usually assumed to be an order of magnitude bigger than the hopping integral $t$,
the relevant AFM coupling between nearest neighbor sites $J=4t^2/U$  is about
tenths of $t$.

Much less is known about the $t$--$J$ model in the small--$J$ limit, i.e.,
the large--$U$ limit of the Hubbard model. At half filling the infinite--$U$
Hubbard model has AFM ground state. However, in 1965 Nagaoka proved a theorem\cite{nagaoka} which
states that when exactly one hole is introduced the ground state becomes FM. In the infinite--$U$
limit the ground state of the half filled Hubbard model is macroscopically degenerate. When a single hole
is introduced this degeneracy is lifted, since it is energetically favorable for the hole to move in a
background of fully aligned spins.
A simple proof of Nagaoka's theorem was later given by Tasaki,\cite{tasaki}
who also showed that additional density-dependent interactions do not alter this result.

The Nagaoka theorem
is one of very few rigorous results concerning strongly correlated electronic systems. However,
it does not say anything about the case of a finite density of
holes as well as finite AFM interaction. The question of a character of the leading instability 
of the Nagaoka state with respect to the AFM exchange term or finite density of holes has 
attracted much interest. 
For finite $J$ ($J>0$) and/or finite doping the ground state is determined by the competition between
antiferromagnetism favored by the exchange interaction and Nagaoka's ferromagnetism favored by the kinetic
energy. This competition presumably drives the system into two phases, a hole-rich FM region
where the kinetic energy is minimized and a region with localized electrons characterized by AFM
order. 
Generally, the $t$--$J$ model may display different types of phase separation depending on the 
value of $J/t$ and 
the actual state of affairs is far from being clear.\cite{ekl,hellberg,tae,riera,putikka,shih,dagotto,
ph_sep,ivanov,batista} It was demonstrated in Refs. \onlinecite{ekl,hellberg,tae} that phase separation takes
place in the $t$--$J$ model for all values of $J$. Other authors\cite{riera,putikka,shih,dagotto} find 
phase separation
only for large $J$. It is very difficult to establish unambiguously the presence or absence of phase separation
in the small--$J$ limit of the $t$--$J$ model. The main reason is that a high-accuracy, unbiased calculation
of the ground state energy as a function of the hole density is required to assess the
competition between the interaction and kinetic energies. The spatial inhomogeneity in the
phase separated state makes analytical approaches rather involved. On the other hand, since the 
FM bubbles are relatively large,\cite{eisenberg} it is difficult
to apply numerical approaches like the quantum Monte Carlo method or exact diagonalization. In this situation,
a computational method that is able to tackle a system sufficiently large to describe the
spatially separated state in the small--$J$ limit of the
$t$--$J$ model
is required. 

In this paper, we demonstrate that Monte Carlo simulations for the recently proposed Ising
version of the $t$--$J$ model\cite{mmfk} provide sound and reliable results in this limiting case.
We employ this approach to study the formation of a bubble of the FM phase when holes are
introduced into an AFM background.
%
Namely, we investigate the so--called Nagaoka polaron which sets in for vanishing
doping and $J/t \ll 1$.\cite{white}
The Nagaoka polaron represents an intermediate state between the homogeneous FM Nagaoka
ground state for $J=0$ and the standard spin polaron for $J>0.2$. 
Numerical studies of the Nagaoka polaron are difficult
because of its large spatial dimensions: for $J \rightarrow 0$ its radius diverges as $J^{-1/4}$.

In this paper we formulate an effective description of the Nagaoka regime,
which is based on the recently proposed Ising version of the $t$--$J$  model.\cite{mmfk}
In the subsequent sections we recall the basic properties of this model
and explain why it gives the same physical picture of the Nagaoka regime as the standard
isotropic $t$--$J$  model. These qualitative arguments are accompanied by quantitative comparison
with the numerical results for the isotropic model.
For the reader convenience we first recall density matrix renormalization group (DMRG) studies\cite{white} on
the Nagaoka {\em polaron}. We then present new results for the Nagaoka {\em bipolaron}, studied in the
$t$--$J$ model by means of exact diagonalization in the limited functional space (EDLFS).\cite{janez1}

Numerical calculations within the Ising version  are by far less demanding hence much bigger clusters
and/or much larger doping  become accessible.  After having successfully tested the single and two--holes cases,
we investigate the ground state properties of the  $t$--$J$ model
($J/t \ll 1$) doped with several holes. In particular, we show that all holes are confined in a single FM polaron.
We discuss how its energy and spatial dimensions depend on the number of holes. For low hole densities, our results provide an evidence that the leading instability of the
FM Nagaoka state is a phase separation rather that a single spin flip.

\section{the Ising $t$--$J$ model}

We start with the $t$--$J$ Hamiltonian on a square lattice \cite{spalek}
\begin{equation}
H_{t-J}=-\sum_{ij\sigma} t_{ij} \tilde{c}_{i\sigma}^{\dagger}
\tilde{c}_{j\sigma}+ J\sum_{\langle ij\rangle} \left(\bm Q_i \bm Q_j -
\frac{1}{4}\tilde{n}_i\tilde{n}_j\right),\label{2.1}
\end{equation}
where
$\tilde{c}_{i\sigma}=Pc_{i\sigma}P=c_{i\sigma}(1-n_{i,-\sigma})$ is
the projected electron operator (to exclude the on-site double
occupancy), $\bm
Q_i=\sum_{\sigma,\sigma'}{c}_{i\sigma}^{\dagger}\bm\tau_{\sigma\sigma'}{c}_{i\sigma'},
\,\bm\tau^2=3/4, $ is the electron spin operator and $\tilde
n_i=Pn_iP=n_{i\uparrow}+n_{i\downarrow}-2n_{i\uparrow}n_{i\downarrow}$
is the projected electron number operator.
Hamiltonian~(\ref{2.1}) contains a kinetic term with the hopping
integrals $t_{ij}$  and a potential $J$ describing the strength of
the nearest neighbor spin exchange interaction. At every lattice
site the Gutzwiller projection operator
$P=\prod_i(1-n_{i\sigma}n_{i-\sigma})$ projects out the doubly
occupied states. Physically this modification of the
original Hilbert space results in strong electron correlation
effects.

At the low-energy scale of order $J\,(\ll t)$, it is reasonable to consider the background
spin configuration to be nearly frozen with respect to the hole dynamics.
In this case the properties of the low-energy quasiparticle excitations in the $t$--$J$ model
are at least qualitatively similar to those in the anisotropic $t$--$J_z$ model,
in which the spin-flip part of the Heisenberg interaction is dropped:
\begin{equation}
H_{t-J_z}=-\sum_{ij\sigma} t_{ij} \tilde{c}_{i\sigma}^{\dagger}
\tilde{c}_{j\sigma}+ J_z\sum_{\langle ij\rangle} \left( Q^z_i Q^z_j -
\frac{1}{4}\tilde{n}_i\tilde{n}_j\right),\label{z1}
\end{equation}
The global continuous spin
SU(2) symmetry of the $t$--$J$ model now reduces to the global
discrete Z$_2$ symmetry of the $t$--$J_z$ model.
Although $Q^z_i Q^z_j$ interaction possesses discrete Z$_2$ symmetry, the
original SU(2) symmetry of all other terms of the Hamiltonian is preserved.
Therefore, the symmetry of the $t$--$J_z$ model depends
on whether  $J_z$ is zero or finite. Namely, for $J_z =0$ the SU(2) symmetry
is restored again.
Although this model is more amenable to numerical calculations, again only rather small lattice clusters
are allowed.

One may hope that the full Ising version of the $t$--$J$ model in which the $t$-term also possesses the global
discrete $Z_2$ spin symmetry results in a more tractable though still nontrivial model.
It by definition has the global
discrete $Z_2$ symmetry, regardless of the values of the incoming parameters. This implies that
the resulting system can be thought of as a doped classical Ising model.
However, it is not clear how such a model can be derived directly from (\ref{2.1}),
since the projected electron operators $\tilde{c}_{i\sigma}$ transform themselves in the
fundamental representation of SU(2). To overcome this problem, the recently proposed spin-dopon
representation of the projected electron operators can be used.\cite{mmfk}

The idea behind that approach is to assign fermion operators $d_{i\sigma}$ to
doped carriers (holes, for example) rather than to the lattice electrons.
The enlarged on-site Hilbert space is spanned by the state vectors $|\sigma a\rangle,$ with
$\sigma=\Uparrow,\Downarrow$ labeling the 2D Hilbert space of the lattice spin,
$\bm{S}_i$, and $a=0,\uparrow,\downarrow,\uparrow\downarrow$ labeling the $4D$ on-site
dopon Hilbert space. The physical space consists of the spin-up $|\Uparrow 0\rangle_i$
spin--down $|\Downarrow 0\rangle_i$ and spinless vacancy $(|\Uparrow \downarrow\rangle_i
- |\Downarrow \uparrow\rangle_i)/\sqrt{2}$ states.\cite{ribeiro} The remaining unphysical
states are removed by the constraint \cite{fku}
\begin{eqnarray}
\bm S_i\bm{M}_i +\frac{3}{4}n^d_i =0 \label{2.3},
\end{eqnarray}
where $\bm M_i=\sum_{\sigma,\sigma'}{d}_{i\sigma}^{\dagger} \bm\tau_{\sigma\sigma'}{d}_{i\sigma'}$
stands for the dopon spin operator so that
\begin{equation}
\bm{Q}_i=\bm{S}_i+\bm{M}_i.
\label{Q}\end{equation}
The physical electron operator $\tilde{c}_{i\sigma}$ is then expressed in terms of the
spin and dopon operators projected onto the physical subspace singled out by Eq. (\ref{2.3}).

In view of the relation
$(S^{\alpha})^2=1/4$, the constraint (\ref{2.3}) can equivalently be written
in the form
\begin{eqnarray}
\sum_{\alpha=x,y,z} S^{\alpha}_i{M}^{\alpha}_i +n^d_i\!\!\!\sum_{\alpha=x,y,z}(S^{\alpha}_i)^2 =0
\label{2.3*}.
\end{eqnarray}
Within the full Ising $t$--$J$ model, one should have $Q_i^{\pm}=Q_i^{x}\pm Q_i^{y}\equiv 0.$
In view of Eq. (\ref{Q}), this requires $S_i^{\pm}=M_i^{\pm}=0$. To explicitly derive the
Ising $t$--$J$ model, we therefore project the dopon operators onto the Hilbert space
singled out by the local constraint
\begin{equation}
S^z_iM^z_i+ \frac{1}{4}n^d_i=0.
\label{2.4}\end{equation}
which can be thought of as an "Ising" form of Eq.(\ref{2.3}). It represents the Z$_2$ singlet
under the $Q^z_i\to \pm Q^z_i$ transformations $\in  Z_2\subset SU(2)$.
The physical electron projected operators reduce to the Z$_2$ spinors:
\begin{eqnarray}
\tilde c_{i\downarrow}&=&{\cal P}^{\rm ph}_i d_{i\uparrow}^{\dagger}{\cal P}^{\rm ph}_i
=\left(\frac{1}{2}-S^z_i\right) d_{i\uparrow}^{\dagger},\label{2.5}\end{eqnarray}
\begin{eqnarray}
\tilde
c_{i \uparrow} &=& {\cal P}^{\rm ph}_i d_{i \downarrow}^{\dagger}{\cal P}^{\rm ph}_i
=\left(\frac{1}{2}+S^z_i\right)d_{i\downarrow}^{\dagger},
\label{2.5*}
\end{eqnarray}
with the projection operator
${\cal P}^{\rm ph}_i=1-(2S^z_iM^z_i+n^d_i/2)$.
It can readily be checked that
$$ Q_i^z=\frac{1}{2}(\tilde {c_{i \uparrow}}^{\dagger}\tilde {c_{i \uparrow}}
-\tilde c_{i\downarrow}^{\dagger}\tilde c_{i\downarrow})
=S_i^z+M_i^z,$$
$$Q^{+}_i=(Q^{-}_i)^{\dagger} =\tilde c^{\dagger}_{i\uparrow}\tilde c_{i\downarrow}\equiv 0,$$
as desired: the transverse components of the electron spin operators no longer appear
in the theory.

The underlying onsite Hilbert space rearranges itself in the following way.
The operators (\ref{2.5})  act on the Hilbert space ${\cal
H}_{\downarrow}=\left\{|\Downarrow,0\rangle,\,|\Downarrow,\uparrow\rangle\right\}$.
These operators do not mix up any other states. Operator
$\tilde c_{i\downarrow}$ destroys the spin-down electron and creates a
vacancy. This vacancy is described by the state
$|\Downarrow,\uparrow\rangle$. The similar  consideration holds for
the $\tilde c_{i\uparrow}$ operators. Now, however, the vacancy is
described by the state $|\Uparrow,\downarrow\rangle$. Those two vacancy
states are related by the $Z_2$ transformation.
The operator
$(Q^z_i)^2=\frac{1}{4}(1-n^d_i)$ produces zero upon acting on the both.
The physical Hilbert state is therefore a direct sum ${\cal
H}_{\rm ph}={\cal H}_{\uparrow}\oplus {\cal H}_{\downarrow}.$  Under the
Z$_2$ transformation $(\uparrow\leftrightarrow\downarrow,\, S^z_i\to
-S^z_i)$ we get ${\cal H}_{\uparrow}\leftrightarrow{\cal
H}_{\downarrow}$, which results in ${\cal H}_{\rm ph}\to {\cal H}_{\rm ph}$.
In the isotropic $t$--$J$ model these two 2D spaces
merge into a 3D SU(2) invariant physical space, where the vacancy is
just an antisymmetric linear combination given by the
SU(2) spin singlet, $(|\Uparrow \downarrow\rangle_i
- |\Downarrow \uparrow\rangle_i)/\sqrt{2}$.
The symmetric combination splits off,
since it represents an unphysical spin-triplet state.

As a result, one arrives at the
representation (\ref{z1}) in which, however, the electron projection
operators are given by Eqs. (\ref{2.5}-\ref{2.5*}). All the parts of this Hamiltonian
possess the global discrete Z$_2$ symmetry
whereas the global continuous SU(2) symmetry is completely lost.
Close to half--filling,
the Ising $t$--$J$ Hamiltonian takes on the form,\cite{mmfk}
\begin{eqnarray}
H^{Ising}_{t-J} &=& \sum_{ij\sigma}t_{ij}
d_{i\sigma}^{\dagger} d_{j\sigma}
+  J\sum_{\langle ij\rangle}\left[\left(S^z_i S^z_j-\frac{1}{4}\right)  \right.
\nonumber \\
 && \left. +  S_i^zM^z_j +S_j^zM^z_i \right],
\label{2.6}
\end{eqnarray}
which is to be accompanied by the constraint (\ref{2.4}).
The magnetic $M_i^zM^z_j$ term has been dropped
as being small of order $\delta^2$
in the limit $\delta:=\langle n_d\rangle \ll 1$.

In practical calculations, we find convenient
to implement the constraint (\ref{2.4})
with the help of a Lagrange multiplier.
Since $S^z_iM^z_i+ n^d_i/4\ge 0$,
the global Lagrange multiplier
\begin{equation}
\lambda\sum_i\left(S^z_iM^z_i+ \frac{1}{4}n^d_i\right)
\label{cons}
\end{equation}
enforces the constraint (\ref{2.4}) locally. The unphysical
occupancy of an arbitrary site  would enhance the total energy by $\lambda\to +\infty$.
Therefore, all unphysical on-site states are
automatically eliminated, so that the local constraint is
taken into account rigorously.

The main difference between the $t$--$J_z$ and Ising $t$--$J$ models
originates from the different symmetries of the hopping terms as discussed in.\cite{mmfk}
The one--hole energy
obtained for the isotropic $t$--$J$ model
is shown to be in between the results obtained for the $t$--$J_z$ and Ising models.
In the regime $J\ll t$, the differences between $t$--$J$ and $t$--$J_z$ models are comparable
to those between $t$--$J$ and Ising models.

\subsection{Monte Carlo approach}
The Hamiltonian (\ref{2.6}) together with the constraint (\ref{cons}) describe a system that
contains both classical ($S^z_i$) and quantum ($d_{i\sigma}$) degrees of freedom.
However, there is no direct interaction between the quantum particles. Therefore, the Ising $t$--$J$
model is closely related to a (multicomponent) Falicov--Kimball model and we can apply an efficient
Monte Carlo (MC) approach derived exclusively for the latter model.\cite{mmkc}
This numerical approach can be applied neither to the standard $t$--$J$ model nor to the $t$--$J_z$ one.
In the latter case one should use the SU(2) invariant constraint (\ref{2.3}) which involves $S^{\pm}$ operators.

The classical MC method has already
been applied to the Ising $t$--$J$ model in order to study dynamics of holes and destruction of the
AFM order with increasing hole concentration.\cite{mmfk} Here, however, we are not
interested in the thermodynamics of the system, but rather in its {\em ground state} properties.
Therefore, the Metropolis algorithm is used for simulated annealing.\cite{kirkpatrick} And once again, since
simulations are performed by random walk through the configuration space of the classical variables,
there is no need for quantum annealing and relatively large lattices can be studied. The size of
the lattice for a given $J/t$ and a given number of holes is adjusted so that the size of the polaron
be always significantly smaller. Since the holes can propagate only within the FM region,
the finite size effects become negligible. Most of the calculations are carried out on
20$\times$20 and 30$\times$30 lattices, but in some cases larger clusters, up to 50$\times$50 are necessary 
as well. From now on, a value of the nearest neighbor hopping amplitude is used
as the energy unit ($t_{ij}=t=1$ for $i$ and $j$ being nearest neighbors and $t_{ij}=0$  otherwise).

\section{Test of a single polaron problem}

Neglecting the spin--flip term is generally  a crude approximation, hence
the standard  and the Ising $t$--$J$ models describe
very different systems. However, we will show that for vanishing doping
and small $J$ both approaches give rise to the same physical picture of the Nagaoka
polaron.

The density matrix renormalization group (DMRG) studies\cite{white} of the
2D $t$--$J$ model have shown that for $J <0.03$ the Nagaoka polaron is indeed stable.
Its size and energy can be determined by balancing
the kinetic energy of a hole freely propagating within a FM bubble
against the magnetic energy of the FM bubble relative to the energy of the N\'eel state.
Minimizing the sum of these two energies one easily finds expressions for the
radius $R$ and energy $E$ of the Nagaoka polaron (see Ref. \onlinecite{white}):
\begin{equation}
R \simeq 1.12 J^{-1/4}, \quad \quad E \simeq -4+9.2\sqrt{J},
\label{white}
\end{equation}
which for $J<0.03$ accurately fit the DMRG data.

\begin{figure}
\includegraphics[width=0.48\textwidth]{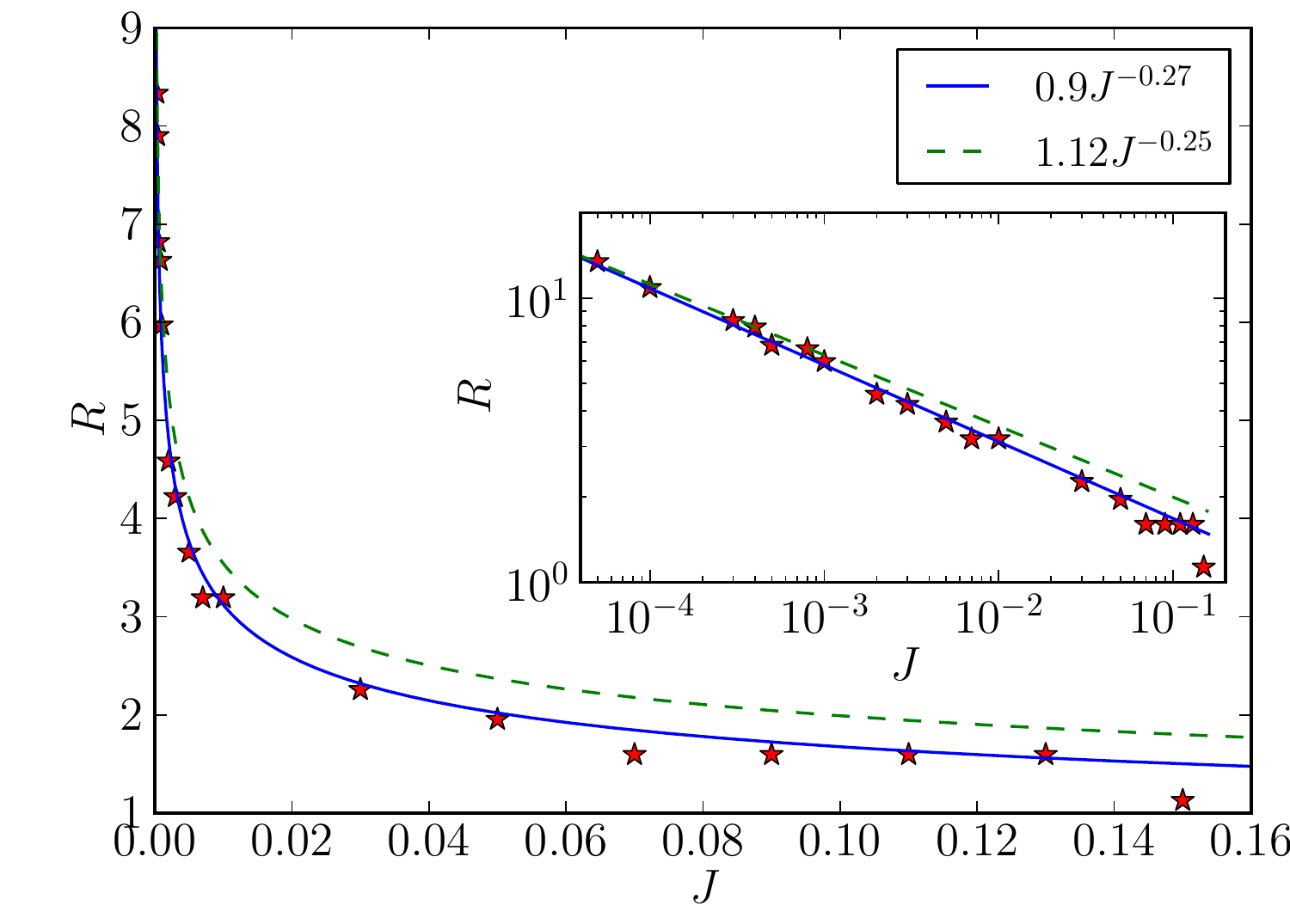}
\caption{(Color online) $J$--dependence of  the radius of the Nagaoka polaron
formed by a single hole. Points show results from the MC calculations for the Ising $t$--$J$ model,
continuous line is the power--law fit and the dashed line shows the dependence described by
Eq.~(\ref{white}). Inset shows the same results but on the log--log scale.}
\label{fig1}
\end{figure}

\begin{figure}
\includegraphics[width=0.45\textwidth]{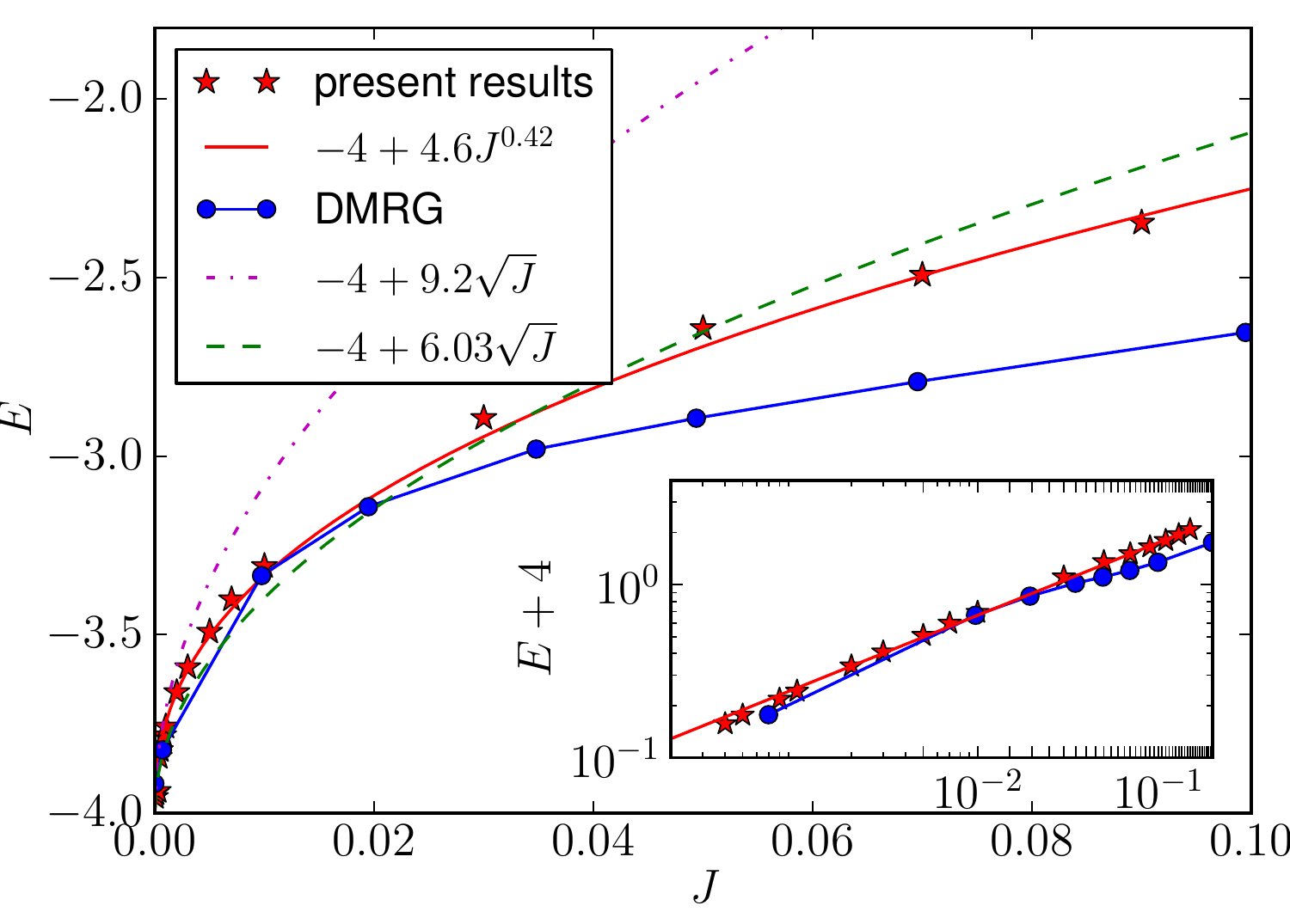}
\caption{(Color online) The same as in Fig. \ref{fig1} but for the energy of a single Nagaoka
polaron relatively to the energy of the homogeneous N\'eel state. The energy is compared with DMRG
results for the $t$--$J$ model taken from Fig. 4 of Ref. \onlinecite{white}. The other lines correspond
to Eq.~(\ref{white}) for the isotropic $t$--$J$  (dot-dashed violet line) and a similar expression for the
$t$--$J_z$ model from Ref. \onlinecite{white} (dashed green line). The continuous line shows a power-law
fit to the present results.}
\label{fig2}
\end{figure}

The Ising $t$--$J$ model displays essentially the same physics.
The constraint (\ref{cons}) allows for propagation of holes only within
{\em  static FM} bubbles where $S^z_j=- M^z_j$ and formation of these bubbles takes place at the expense
of the exchange interaction, which favors AFM alignment  of $S^z_i$. In both models
the magnetic energy of a bubble is qualitatively similar and is proportional to $J$ multiplied by the number
of FM bonds. Quantitative differences should arise from different energies per bond of the N\'eel ground state of the undoped systems.
Figs. \ref{fig1} and \ref{fig2} show  quantitative comparison of both models. Here, we show the ground
state properties of a single Nagaoka polaron in the Ising $t$--$J$ model for a $50 \times 50$ system
with periodic boundary conditions and $\lambda=300$. The radius and energy of the Nagaoka polaron  have been compared with expressions  (\ref{white}) as well as with the bare DMRG data
for the SU(2) $t$--$J$ model taken from Ref. \onlinecite{white}. The overall agreement is clearly visible.

To conclude this section, we notice that the Ising $t$--$J$ model provides a simple and
reasonably accurate description of a single Nagaoka polarons for $J \ll 1$ simply because
the  spin--flip term becomes irrelevant inside a sufficiently large FM bubble which confines the movement of holes.

\section{The Nagaoka bipolaron problem}

In the preceding section we have successfully tested the case of a single carrier doped into Mott insulator.
Since our aim is to study a system at finite doping such a test might still be insufficient. Therefore,
we investigate also the system doped with two holes. On the one hand it is a first nontrivial step
towards understanding the spatial hole distribution  in doped systems (uniform versus inhomogeneous).
On the other hand,  due to large spatial dimensions of the Nagaoka polaron, it is already
a challenging problem for fully quantum numerical approaches.

To investigate the Nagaoka bipolaron in the isotropic $t$--$J$ model, we employ the EDLFS method. This method
describes properties of carrier/carriers doped into a planar ordered antiferromagnet.\cite{janez1}
One starts from a translationally invariant state of two carriers in the N\' eel background
\begin{equation}
\vert \phi_{0}{\rangle}_p = \sum_{\boldsymbol{\gamma}}(-1)^{M(\boldsymbol{\gamma})}
c_0c_{\boldsymbol{\gamma}}\vert {\rm Neel }\rangle,
\end{equation}
where the sum runs over two  nearest neighbors to site 0 and  $M(\boldsymbol{\gamma})$  sets the appropriate sign to generate   $p_{x(y)}$-wave symmetry.
The kinetic part $H_k$ as well as the off--diagonal spin--flip part $\tilde{H}_J$ of the Hamiltonian  (\ref{2.1})
are applied
up to $N_h$ times generating the basis vectors:
\begin{equation}
\left\{|\phi_{l}^{n_h} \rangle \right\}=[H_k + \tilde{H}_J]^{n_h}
|\phi_0 \rangle_p, \quad n_h=0,...,N_h.
\label{EDLS}
\end{equation}
Then, the ground state $|\Psi_0\rangle$ is calculated within the limited functional space by means of the
Lancz\"os method. The advantage of EDLFS over the standard exact diagonalization (ED) approach follows
from systematic generation of selected states which contain spin excitations in the vicinity of the
carriers. It  enables investigation  of much larger systems, which is particularly important in the
Nagaoka regime.  We apply this method and study the average distance $D$ between two holes
in the $t$--$J$ model.

From the construction of EDLFS follows that $N_h$ determines maximum distance between
holes and spin excitations as well as the maximum accessible $D$.
Therefore, all characteristic length--scales of investigated problem should be smaller
than a certain $\xi(N_h)$. Since the successive application of the nearest neighbor hopping [see Eq. (\ref{EDLS})]
closely resembles the random walking process we expect that $\xi (N_h) \propto \sqrt{N_h}$.
The numerical complexity of the two--holes problem allows us to study $N_h$ up to 12.
Therefore, for small $J$ we carry out a finite size scaling with respect to the size of the Hilbert space
and study $D(J) \equiv D(J, N_h \rightarrow \infty)$.
The finite size effects should vanish when
$\xi (N_h)/D(J) \gg 1 $ or equivalently $N_h/D^2(J) \gg 1 $. We have
found that this vanishing can universally be described by the Gaussian:
\begin{equation}
D(J)-D(J,N_h)=\alpha \exp\left\{-\beta\left[\frac{N_h}{D^2(J)}\right]^2\right\},
\label{extr}
\end{equation}
where parameters $\alpha$ and $\beta$ are independent of $J$.
Note that for a given $J$ fitting of $D(J,N_h)$ involves only a single free parameter
$D(J)$ which simultaneously represents the extrapolated distance between two holes.
\begin{figure}[h]
\includegraphics[width=0.53\textwidth]{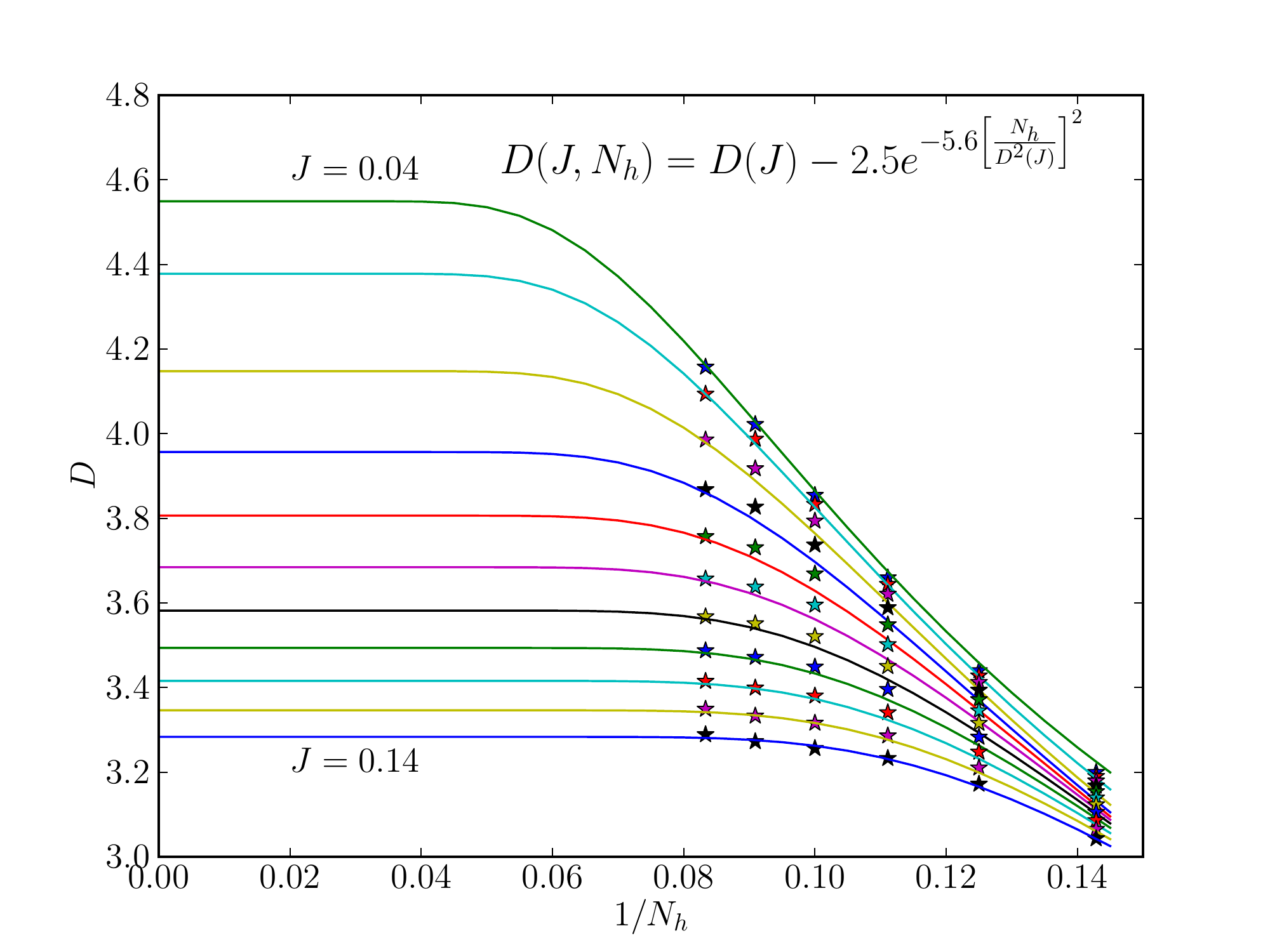}
\caption{(Color online) Distance between two holes $D$ in the isotropic $t$--$J$ obtained in a functional
Hilbert space generated for finite $N_h$.
Numerical results (points) are fitted and extrapolated according to  Eq. (\ref{extr}).}
\label{extrap}
\end{figure}
Fig.~\ref{extrap} shows the extrapolation, while the resulting $D(J)$ is shown in  Fig.~\ref{fit_extrap}.
\begin{figure}[h]
\includegraphics[width=0.53\textwidth]{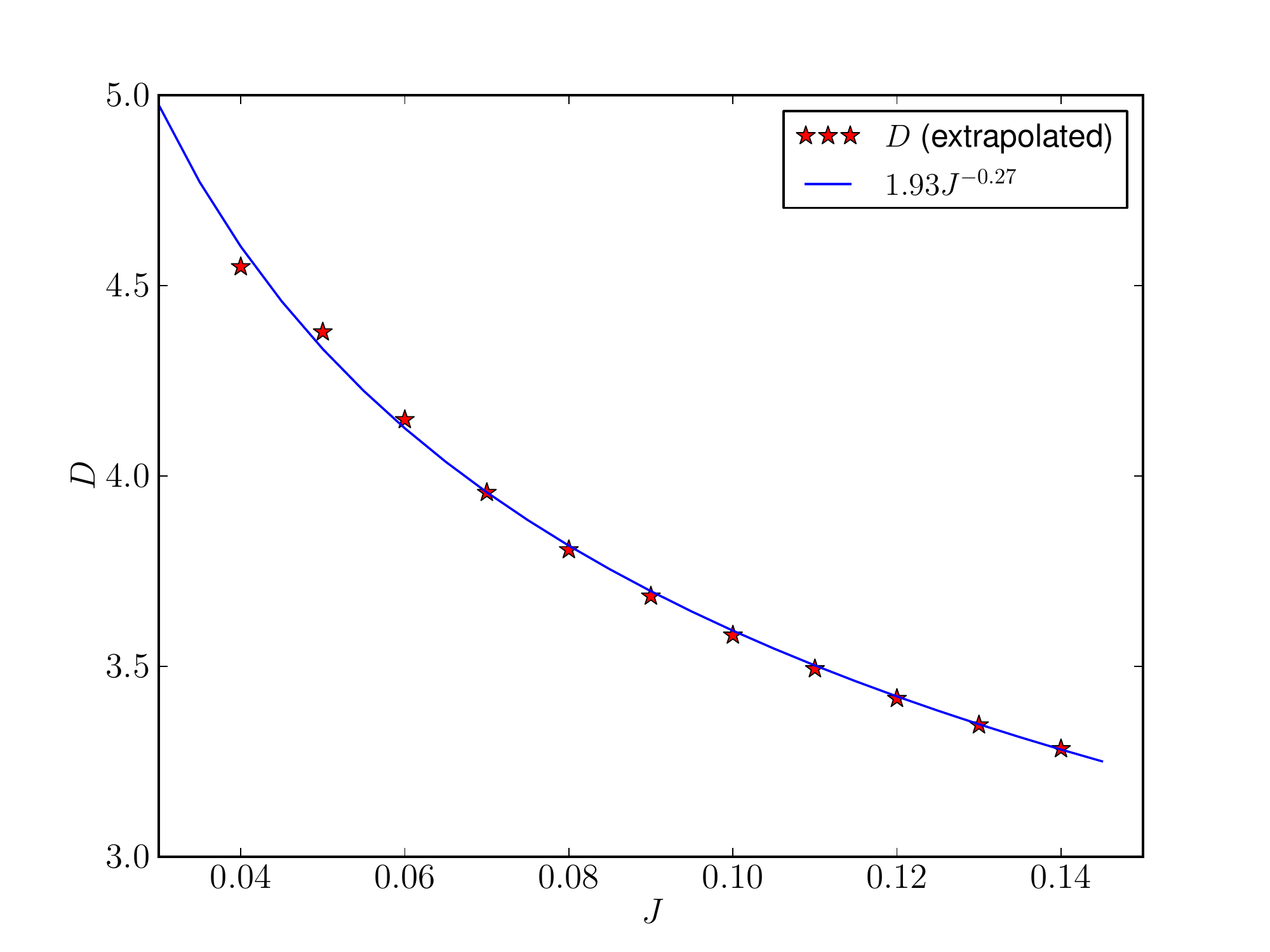}
\caption{(Color online) Extrapolated distance between two holes D(J) for the isotropic $t$--$J$ model.}
\label{fit_extrap}
\end{figure}
Finally, the quality and the universality of the finite--size scaling is directly shown  in  Fig.~\ref{scaling}.
\begin{figure}[h]
\includegraphics[width=0.53\textwidth]{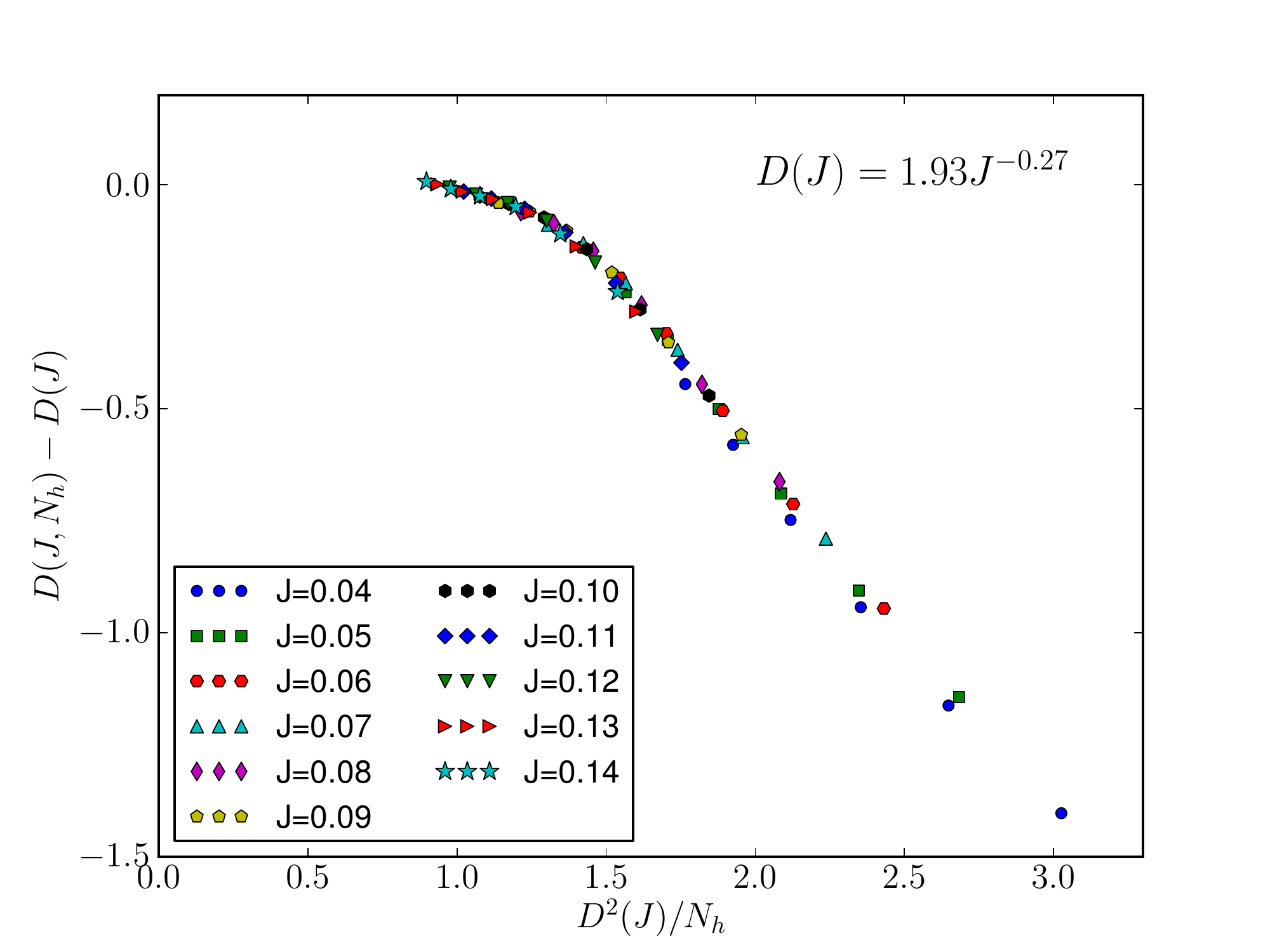}
\caption{(Color online) The same data as in Fig. \ref{extrap}, but rescaled according to Eq. (\ref{extr}).}
\label{scaling}
\end{figure}
One can clearly see that all data for
$J=0.04,\ldots, 0.14$ merge into a single curve, which strongly supports the assumptions behind
Eq. (\ref{extr}). This also demonstrates a self-consistency of the approach.

These results provide solution of the long--standing open problem, i.e., whether two holes in the $t$--$J$ model form a bound state in the small $J$ limit.~\cite{twoholes,leung02}
Since $D(J) > 3.5$ for $J < 0.10$, the problem can hardly be solved by exact diagonalization on 32--site cluster~\cite{leung02} where the maximal possible distance between two holes is four lattice spacings.
Note that the symmetry of the bound state is $p$--wave for $J \lesssim 0.15$.
Therefore, the results using the EDLFS method indicate that two holes in the single band $t$--$J$ model are always bound, which may not necessarily be the case in more general models describing the ${\rm CuO_2}$ plane.~\cite{lau11}

\begin{figure}[h]
\includegraphics[width=0.47\textwidth]{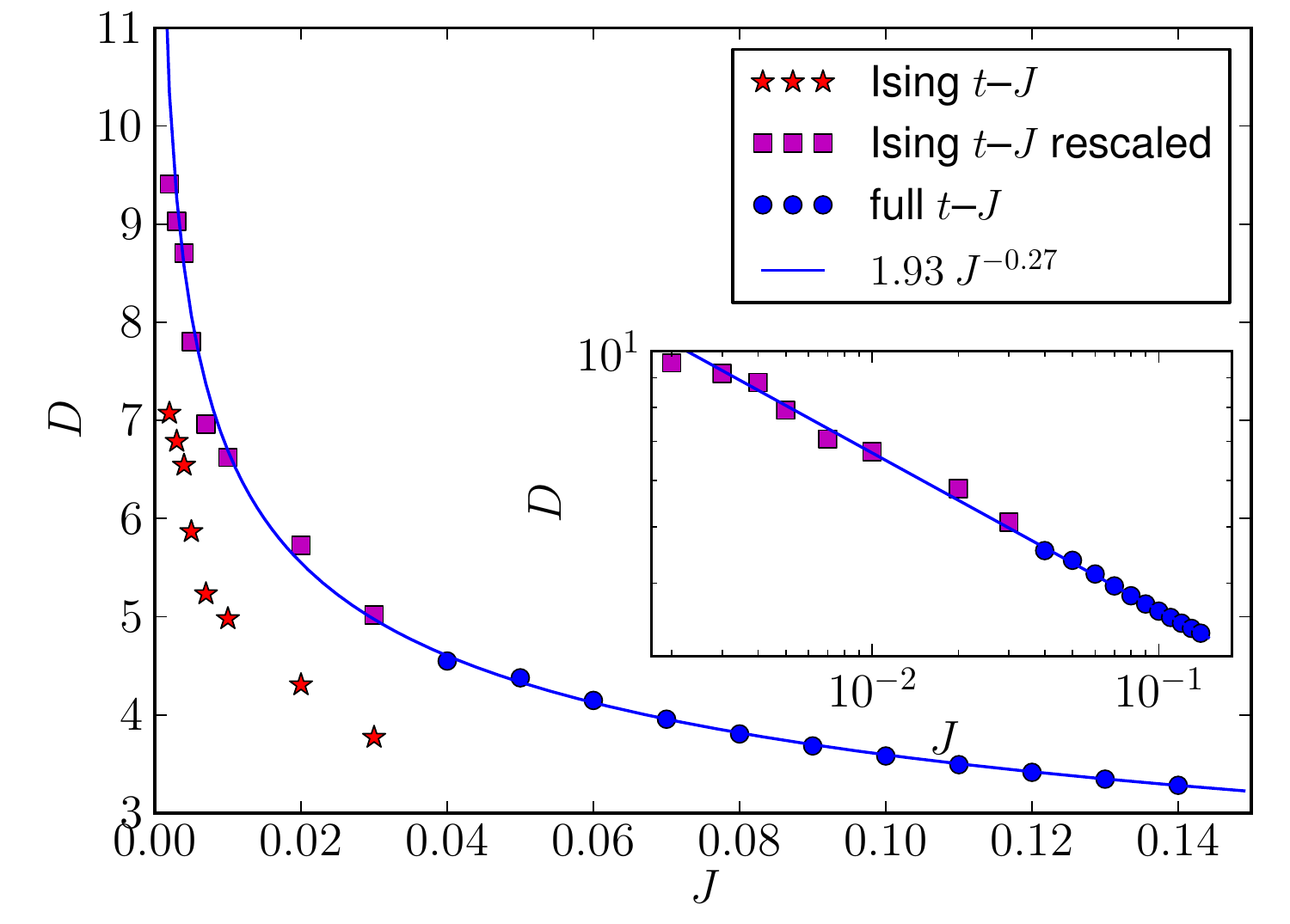}
\caption{(Color online) Comparison of the distance between two holes in the $t$--$J$ (blue circles) and
Ising $t$--$J$ (red stars) models. "Ising $t$--$J$ rescaled" means that all values are multiplied by
1.33. Note that the rescaled values can be described by Eq. (\ref{fitD}), see also the inset where compounded $D(J)$ data  are presented on the log - log scale.
}
\label{compar}
\end{figure}

The most important result shown in Fig. \ref{fit_extrap} concerns the power--law dependence:
\begin{equation}
D(J) \simeq 1.93J^{-0.27},
\label{fitD}
\end{equation}
hence $D(J)$ is roughly proportional to the radius of a single--hole polaron $R(J)$ [see Eq. (\ref{white})].
Already this result suggests that the linear dimensions of polaron and bipolaron are determined
by the same mechanisms, which are properly captured by the Ising $t$--$J$ model. In order to show that
this expectation holds true we have  calculated $D$ in the Ising version of the $t$--$J$ model.
Results are shown as (red) stars in Fig \ref{compar}.
In the same figure results for the $t$--$J$ model are presented as (blue) dots. One can clearly see that the
distance between holes $D$ in the Ising $t$--$J$ model increases with decreasing $J$ slower than in the
full $t$--$J$ model. This result is not a surprise: in the isotropic
$t$--$J$ model (as well as in the $t$--$J_z$ one) a hole is able to enter the AFM
surrounding of the FM bubble. Since the boundary of the FM
bubble is unpenetrable  in the Ising $t$--$J$ model, we expect a smaller average distance between holes in the latter case.
However, it turns out that the
difference has only a quantitative character, i.e., only the coefficient in Eq. (\ref{fitD}) is approx. 33\%
smaller. This agreement shows that these two methods, i.e., the EDLFS method  for the full $t$--$J$ model
and the MC method for the Ising $t$--$J$ model are complementary in a sense: The applicability of
the EDLFS method is limited by the maximum size of the Hilbert space, and since the distance between holes
increases with decreasing $J$, this method cannot be used when $J$ is too small. On the other hand, the
importance of the spin--flip term in the $t$--$J$ Hamiltonian diminishes with decreasing $J$ and the
approximation that leads to the Ising $t$--$J$ model becomes  more reliable in the region where the EDLFS
method cannot be  applied any longer. The main advantage of the Ising $t$--$J$ model is that it can be studied
within the framework of the classical MC method on clusters sufficiently large to describe large
polarons that emerge at small $J$.

\section{FINITE HOLE DENSITY}
Up to this point we have been analyzing one and two holes in the whole system. Since the size of the (bi)polaron and
its energy do not depend on the size of the lattice [provided the lattice is significantly larger than the
(bi)polaron size], these results effectively describe the case of the vanishing density of holes. Then, the
important question arises as to how  the ground state of the Ising $t$--$J$ model evolves when the number of holes
increases. Possible scenarios include phase separation or homogeneous distribution of holes. It is also
possible that a single polaron becomes unstable at some critical value of the hole number
giving way to smaller polarons.

In order to study this problem we calculate the total energy of the system as a function
of the number of holes $E(N)$. Convex $E(N)$ for some $N$ indicates that it is energetically favorable
to split a FM bubble with $N$ holes into smaller bubbles with $M<N$ holes,
provided that $E(M)$ is  concave. If $E(N)$ is convex for
arbitrary $N$, holes will not form polarons with more than one hole. On the other hand, if $E(N)$ is
concave for arbitrary $N$, all holes introduced into the system will gather in a single FM
region. In other words, the phase separation into a hole--rich FM region and an AFM region
without holes takes place.

\begin{figure*}[h]
\includegraphics[width=\textwidth]{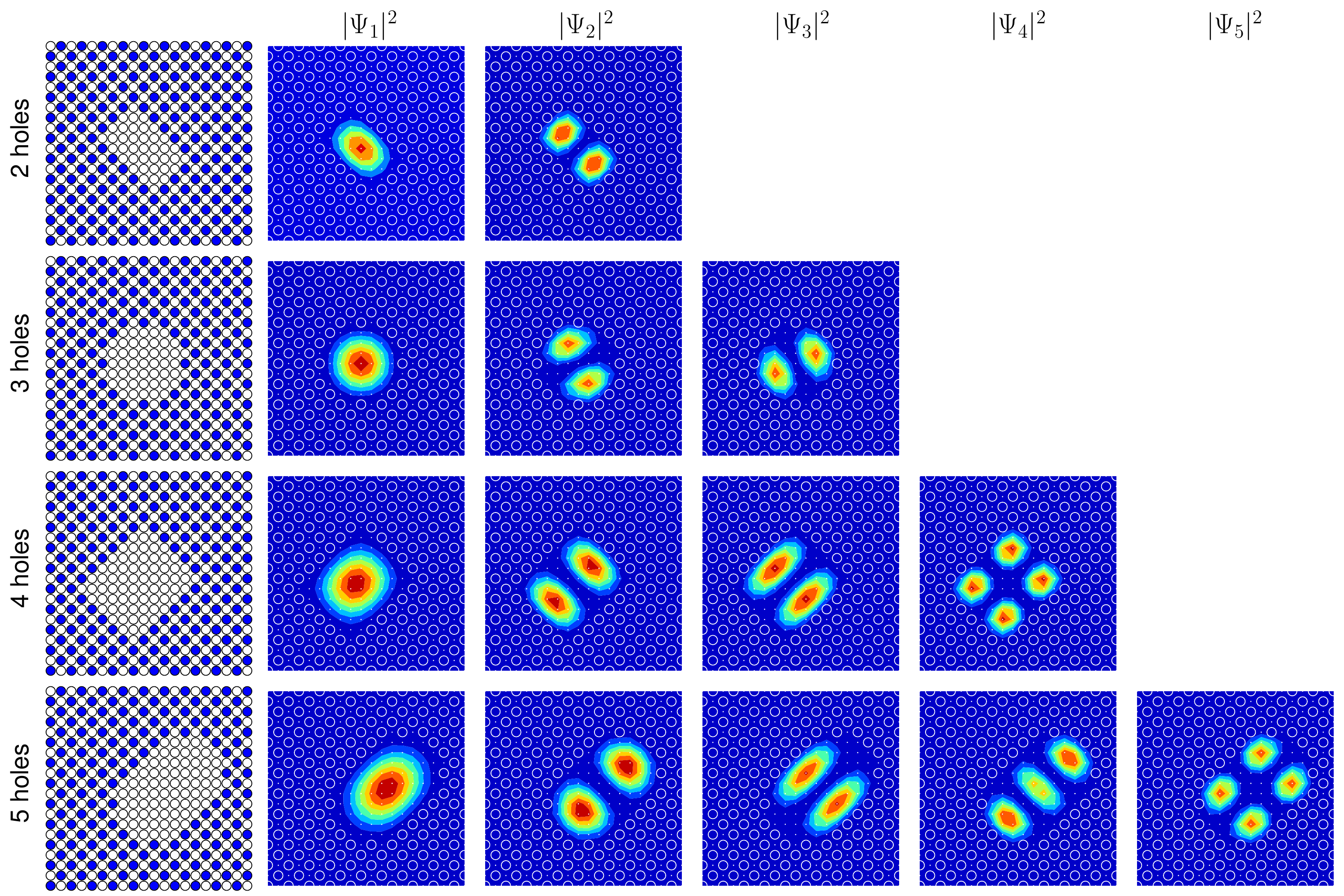}
\caption{(Color online) Leftmost column: shapes of the Nagaoka polarons including from 2 to 5 holes. Filled
circles indicate spin-up lattice sites and empty circles spin-down lattice sites. The rest of the panels
show wave functions of the holes. In all cases J=0.03 was assumed.}
\label{multipolarons}
\end{figure*}

Accurate determination of $E(N)$ is generally a difficult task. For $N=1$ the shape of the polaron is
almost circular, but for higher $N$ the geometry becomes nontrivial.
Fig. \ref{multipolarons} shows the Nagaoka polarons and corresponding hole wave functions for $N=2,\ldots,5$.
The shape  of the polaron follows directly from the spatial structure of the occupied orbitals. The diagonal orientation of almost rectangular polarons minimizes the magnetic energy along the line between
FM and AFM regions.

\begin{figure}[h]
\includegraphics[width=0.53\textwidth]{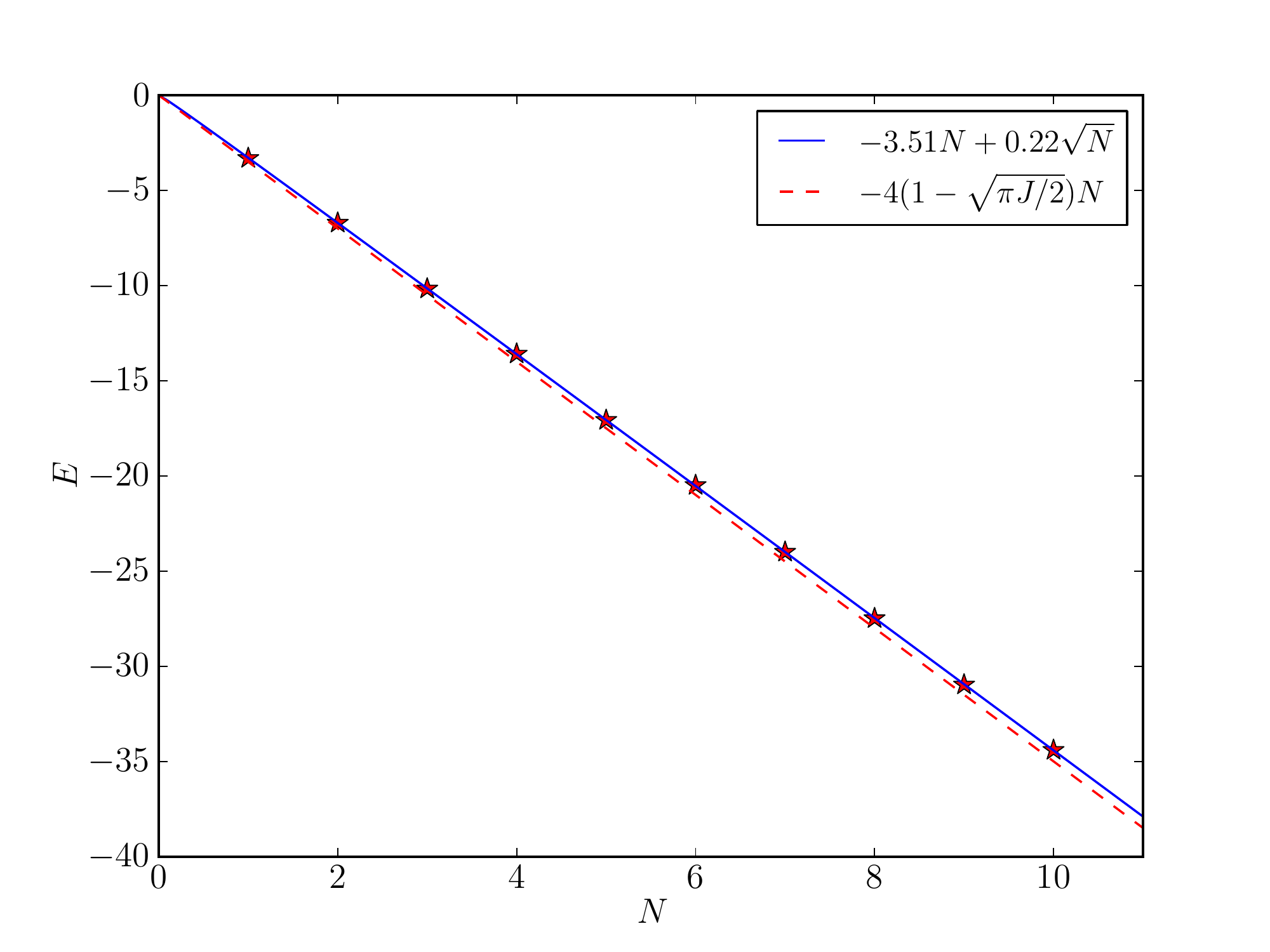}
\caption{(Color online) Energy $E(N)$ of a polaron containing $N$ holes relatively to the energy of the homogeneous
N\' eel state (points) for $J=0.01t.$ The full line represents a fit to the numerical data as given in the legend, the dashed line represents analytical result in Eq.~(\ref{Etot}).}
\label{E_vs_N}
\end{figure}

With increasing  $N$ the size of the FM bubble increases, so a large lattice is necessary to avoid the
finite size effects. The advantage of the Ising version of the $t$--$J$ model
becomes evident,
since it can be reliably investigated on lattices much larger than those accessible to the
fully quantum methods like quantum MC, exact diagonalization or even EDLFS.
Using larger lattices we study polarons containing up to 10 holes.
In Fig. \ref{E_vs_N} we show the polaron energy $E(N)$ as a function of the number of holes $N$
for $J=0.01$.
Studying other values of  $J$ from 0.01 to 0.1 (not shown) we fitted the energy with a function
$E(N)=a N + b\sqrt{N}$. In Sec. V.A. we justify such a form of $E(N)$.  We have found that in all cases $b$ is positive,
which means that $E(N)$ is concave.
This in turn implies that the Ising version of $t$--$J$ model displays a phase separation for all
those values of $J$ at which it is still equivalent to the isotropic $t$--$J$ model.

An important question concerns the fraction of the system occupied by each of the magnetic phases. It can be  answered by comparing the size of the polaron to the size of the whole system.
\begin{figure}[h]
\includegraphics[width=0.50\textwidth]{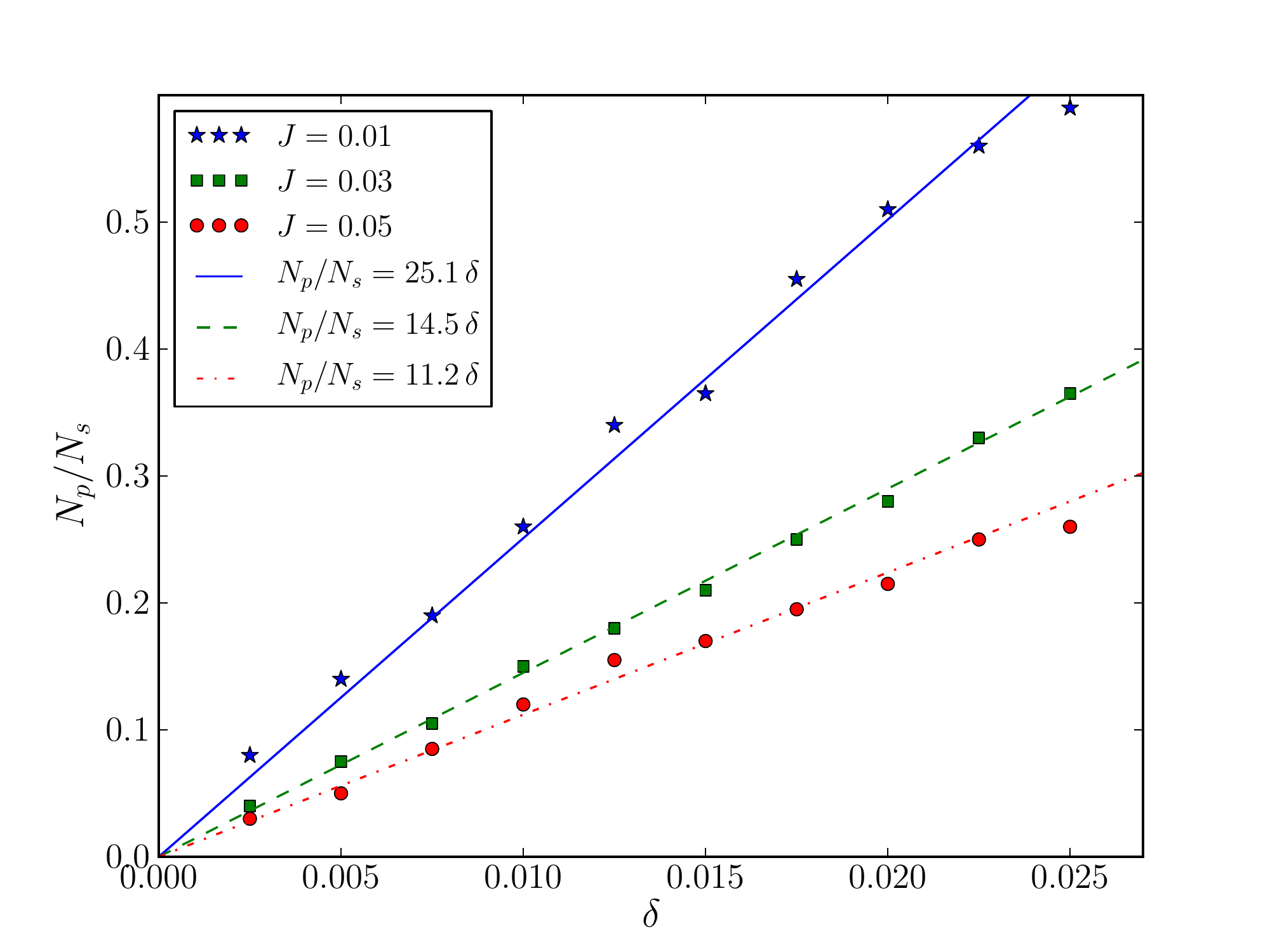}
\caption{(Color online) Fraction of the lattice occupied by the FM polaron as a function
of the concentration of holes for different values of $J$. The straight lines represent 
analytical results given in Eq.~(\ref{ratio}).}
\label{R_vs_N}
\end{figure}
Fig. \ref{R_vs_N} shows
the relation between the fraction of the lattice sites with ferromagnetically aligned spins $N_p/N_s$ and the density of holes $\delta=N/N_s$.
$N_p$ is the number of lattice sites in the polaron and $N_s$ is the size of the lattice.
This dependence can be fitted by a linear function, which extrapolated to $N_p/N_s=1$
gives the threshold value of the hole density $\delta_t$.
If the concentration of holes is close but still smaller
than $\delta_t$, most of the system is occupied by the FM phase,
while the rest forms an AFM island (or islands). Finally,
for concentrations larger than $\delta_t$ the whole system would be in a fully polarized state.
However, the latter regime is probably not accessible by the present approach.
In the Nagaoka regime, the Ising and isotropic $t$--$J$ models give the same results because the
physics of the Nagaoka regime is determined by the competition between the magnetic and kinetic energies.
However, as soon as the FM bubble covers the whole system other mechanisms come into play, e.g.,
a direct hole-hole interaction and/or interference of the carriers paths around loops.
\begin{figure}[h]
\includegraphics[width=0.47\textwidth]{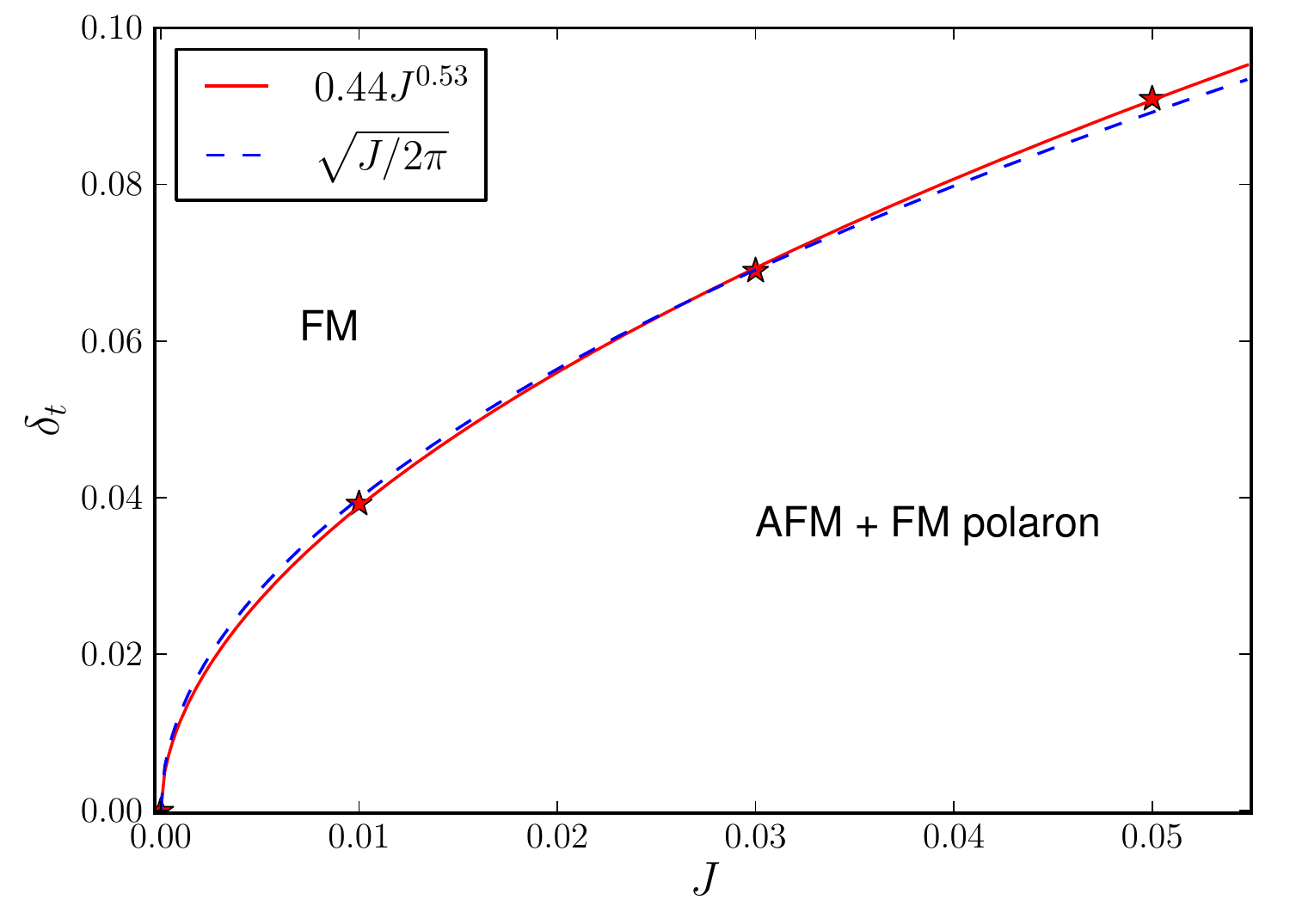}
\caption{(Color online) Critical value of the concentration of holes above which the whole system is
in a fully polarized FM state. The point at $J=0$ is added as a result of the Nagaoka theorem. The dashed line represents analytical result, Eq.~(\ref{deltacr}).}
\label{delta-crit}
\end{figure}
Fig. \ref{delta-crit} shows $\delta_t$ as a function of $J$ (the point $\delta_t=0,\ J=0$ is a result
of the Nagaoka theorem). The obtained square--root dependence between both quantities follows immediately
from Eq. (\ref{white}) and the proportionality $N_p \propto N$ shown in Fig. \ref{R_vs_N}.

\subsection{Analytical approach for finite doping}

We consider the FM polaron of size $N_p$  with  $N$ doped holes that can be treated as spinless noninteracting fermions. The FM polaron is furthermore placed in the hole-depleted N\' eel spin background. We furthermore consider the limit of small hole-density that allows for quadratic expansion of the  single particle kinetic energy:
\begin{equation}
E_{\rm kin}^{(1)}(k) = -2(\cos{k_x}+\cos{k_y})\sim -4 +k^2.
\label{ekin1}
\end{equation}
We obtain the kinetic energy of $N$ holes by integrating Eq.~(\ref{ekin1}) up to $k_F=2\sqrt{\pi N/ N_p}$
\begin{eqnarray}
E_{\rm kin}&=& -4N + 2\pi N^2/N_p.
\end{eqnarray}
We proceed by writing the total energy $E(N)$ as:
\begin{eqnarray}
E(N) &=& E_{\rm kin} + E_{\rm spin},\nonumber \\
E(N) &=& -4N + 2\pi N^2/N_p + N_pJ,
\end{eqnarray}
where the last term represents the magnetic energy of the FM polaron  relative  to the energy of the N\' eel state. After the minimization ${\partial E/ \partial N_p}=0$ we obtain 
\begin{equation}
{N_p\over N_s}  = \sqrt{2\pi\over J}\delta,
\label{ratio}
\end{equation}
representing the ratio between the polaron size $N_p$ and  the total size of the system $N_s$ as a function of hole doping $\delta$. The comparison of Eq.~(\ref{ratio}) with numerical data is shown in Fig.~\ref{R_vs_N}. From
Eq.~(\ref{ratio}) we obtain  as well the critical doping for the transition to the FM state
\begin{equation}
\delta_t = \sqrt{J \over 2\pi},
\label{deltacr}
\end{equation}
shown along the numerical results in Fig.~\ref{delta-crit}.
Notice, that Eq. (\ref{deltacr}) agrees with that derived within the semiclassical calculations of the 2D
isotropic $t$--$J$ model\cite{eisenberg} which
suggest that at small hole concentration and rather weak AFM coupling the
FM Nagaoka state becomes unstable towards a creation of an AFM bubble.
This agreement is quite natural, since the spins are considered to be frozen
in both the classical large-spin limit of the isotropic $t$--$J$ model and
its full Ising version. 
Finally, we obtain the total energy of the system
\begin{equation}
E(N) = -4t\left(1-\sqrt{\pi J\over 2}\right)N.
\label{Etot}
\end{equation}

The comparison with numerical results is shown in  Fig.~\ref{E_vs_N}. 
Here we have neglected the effects along  the line, separating the FM polaron from the N\' eel 
spin background  as well as the dependence of the kinetic energy on the shape of
the bubble.  As a results only the linear term in $E(N)$ is reproduced.
Since the phase separation is determined by the nonlinear part of $E(N)$, 
these effects give rise to small, nevertheless important corrections.
The former one is proportional to the length of the borderline ($\propto \sqrt{N}$) and 
it was the reason for the choice of the fitting function in Fig. \ref{E_vs_N}.

\section{Summary}
The main difficulty in analyzing the $t$--$J$ model in the small--$J$ limit is that a large
size of the lattice is required to correctly describe the dynamics of holes.
This requirement significantly restricts
the applicability of numerical approaches like the quantum MC or exact
diagonalization method.

In the small--$J$ regime, however, the holes are confined in a FM polaron,
so that the spin--flip processes are strongly reduced.
This justifies the applicability of the Ising version of the $t$--$J$
model to study the small--$J$ limit of the original $t$--$J$ model.

For small but finite values of $J$, our results for one and two holes
are in a good agreement with those obtained within the fully quantum approaches
(DMRG, EDLFS). However, we are able to extend our calculations to the regimes
of smaller $J$ and larger number of holes inaccessible by the former methods.
We show that it is energetically favorable
for the system to segregate
into the FM hole--rich phase and hole--depleted AFM phase. The size (surface)
of the FM  bubble depends linearly on the number of holes while its dependence on $J$ is given by
the square--root function. With increasing concentration of holes and/or with decreasing $J$ the size
of the FM polaron increases and eventually for 
$\delta_t\simeq 0.44 \sqrt{J}$
it occupies the whole lattice. 
Our numerical results thus suggest that Nagaoka state breaks down by forming an AFM bubble.
This observation fully agrees with a conjecture discussed earlier within the isotropic $t$--$J$ model.\cite{eisenberg} We, however, expect that
the results obtained for isotropic and Ising $t$--$J$ models start to deviate from each other
when doping becomes larger that the threshold density $\delta_t$ even for
$J \ll 1$.

A rather simple analytic treatment of the holes doped in the FM polaron that is furthermore placed in a N\' eel, hole-depleted spin background, leads to a good agreement with numerical data. Among other results, it  provides a simple expression for the threshold density  
$\delta_t=\sqrt{J/2\pi}$. 
The theory reproduces  only the  linear dependence of the total energy on the number of holes
and does not provide  information on whether the system phase separates.
Corrections 
 (e.g., due to line contributions) are expected to give rise to  a positive $\sqrt{N}$ term in the total energy,  as obtained from the numerical data.

\acknowledgments
M.M.M. acknowledge support from the Foundation for Polish Science under the ``TEAM'' program for the years
2011-2014. M.M. and M.M.M. acknowledge support under Grant No. N N202 052940 from Ministry of Science and Higher Education
(Poland).  J.B. and L.V. acknowledge support under Grant No.  P1-0044 from ARRS (Slovenia). J.B. acknowledges the Gordon Godfrey  bequest of UNSW, Sydney (Australia) where part of this work has been performed.

\end{document}